\documentclass{aa}
\usepackage{graphicx}
\usepackage{natbib}
\usepackage{color}
\usepackage{ulem}
\usepackage{txfonts}
\usepackage[colorlinks=true,linkcolor=blue,citecolor=blue]{hyperref}
\pdfoutput=1
\begin{document}

   \title{Imaging-spectroscopy of a band-split type II solar radio burst with the Murchison Widefield Array}
    \titlerunning{Imaging-spectroscopy of a band-split type II radio burst}
\authorrunning{Bhunia et al.}
   \author{Shilpi Bhunia
          \inst{1},\inst{2} Eoin P. Carley\inst{1}
          ,
          Divya Oberoi\inst{3}
          \and
          Peter T. Gallagher\inst{1}
          }

   \institute{Astronomy \& Astrophysics Section, Dublin Institute for Advanced Studies, Dublin, D02 XF86, Ireland.
         \and
             School of Physics, Trinity College Dublin, College Green, Dublin 2, Ireland.\\ 
            \email{bhunias@tcd.ie}
           \and
          National Centre for Radio Astrophysics, Tata Institute of Fundamental Research, Pune 411007, Maharashtra, India.
             }

   \date{Received ...}

 
\abstract{
Type II solar radio bursts are caused by magnetohydrodynamics (MHD) shocks driven by solar eruptive events such as Coronal Mass Ejections (CMEs). Often both fundamental and harmonic bands of type II bursts are split into sub-bands, generally believed to be coming from upstream and downstream regions of the shock; however this explanation remains unconfirmed. Here we present combined results from imaging analysis of type II radio burst band-splitting and other fine structures, observed by the Murchison Widefield Array (MWA) and extreme ultraviolet observations from Solar Dynamics Observatory (SDO)/Atmospheric Imaging Assembly (AIA) on 2014-Sep-28. The MWA provides imaging-spectroscopy in the range of 80-300\,MHz with a time resolution of 0.5\,s and frequency resolution of 40\,kHz. Our analysis shows that the burst was caused by a piston-driven shock with a driver speed of $\sim$112\,km s$^{-1}$ and shock speed of $\sim$580\,km s$^{-1}$. We provide rare evidence that band-splitting is caused by emission from multiple parts of the shock (as opposed to the upstream/downstream hypothesis). We also examine the small-scale motion of type II fine structure radio sources in MWA images. We suggest that this small-scale motion may arise due to radio propagation effects from coronal turbulence, and not because of the physical motion of the shock location. We present a novel technique that uses imaging spectroscopy to directly determine the effective length scale of turbulent density perturbations, which is found to be 1 - 2\,Mm.
The study of the systematic and small-scale motion of fine structures may therefore provide a measure of turbulence in different regions of the shock and corona.} 

   \keywords{techniques : imaging spectroscopy - Sun: activity- Sun : shock waves- Sun : turbulence}

   \maketitle
%

\section{Introduction}

Type II solar radio bursts are produced by magnetohydrodynamics (MHD) shocks driven by solar eruptive events such as coronal mass ejections (CMEs), flares, and jets \citep{nelson1985,2011Nindos,Gopalswamy,Zucca2018,Maguire_2021,2021Jebaraj, 2021Alissandrakis}. 
The shock waves accelerate electrons, producing Langmuir waves and subsequently plasma emissions at the fundamental ($f_p$) and second harmonic ($2f_p$) of the plasma frequency, given by $f_p \approx 8980\sqrt{n_{e}}$, where $n_e$ is electron number density (cm$^{-3}$). 
Often the dynamic spectra of type IIs exhibit various kinds of fine structures (FS). For example, the fundamental and harmonic bands of type II bursts drift to lower frequencies over time, and can sometimes split into sub-bands, known as `band-splitting'. Type IIs can exhibit FS on a variety of temporal and spectral scales, and altogether such structure can give insight to shock kinematics, geometry, propagation, and turbulence in the solar atmosphere. Such properties are highly important for shock particle acceleration physics. 

 In a study based on 112 metric type II bursts \cite{vrsnak2001} reported only 20\% of type II bursts show the band-splitting phenomena. It is not known in literature why such a small fraction of the type II bursts show such structure. Despite this small fraction, band-splitting shows common features across different events. For example, the relative frequency split between the sub-bands $\triangle f/f$ is found to be within the range of 0.1-0.5 and this value remains constant over the time duration of the event  \citep{vrsnak2001}. 

It is still unclear what causes the band-splitting of the type II bursts,
but there are two widely accepted theories. The first popular mechanism was suggested by \citet{smerd1974,smerd1975} where the high frequency sub-band (HFS) and low frequency sub-band (LFS) of type II are caused by coherent plasma emissions simultaneously coming from the downstream (behind) and upstream (ahead) regions of the {\color{black} same} shock front. This is commonly used to derive shock compression ratio, Mach numbers \citep[e.g.][]{vrsnak2001,Zucca2018,2020Maguire} and the coronal magnetic field strength \citep[e.g.][]{2017Kumari,2019Kumari}
. There have been some imaging studies that have found evidence supporting this interpretation \citep[e.g.][]{zimovets2012,Zucca2018,Chrysaphi2018}. In these studies, band-splitting was most often associated with a CME-driven shock, and the sub-band sources were closely spaced in imaging (\cite{Chrysaphi2018} reported the apparent spatial separation between two sub-band sources at 32 and 40\,MHz for fundamental emission to be $\sim$0.2$\pm$0.05 R\textsubscript{\(\odot\)}).

There have also been results that are not in the favour of the upstream/downstream scenario. \cite{Du_2014} found that the spectral features first appeared in the HFS seconds earlier than in the LFS which according to the upstream/downstream scenario should have happened the other way around, that is the spectral structures should first appear in LFS as the emission should first come from low density upstream (shock ahead) region. 
According to review work by \cite{2011Cairns}, in the downstream region the electron beam distributions are unstable to the production of enhanced Langmuir waves. Also, recent observations \citep{Zucca2018,Magdaleni_2020,2020Morosan,2021Kouloumvakos} of type II dynamic spectra have shown the burst having multiple sub-bands which can not be explained by conventional upstream/downstream scenario.
Hence, an alternative mechanism can be used to explain the band-splitting phenomenon, where spatially separated parts of the shock front produce radio emissions at LFS and HFS \citep{mclean1967}. Until now there have been evidences supporting the upstream/downstream scenario whereas there has been very little evidence for this theory \citep{2015Zimovets}.\par
While type IIs band-splitting is the most common fine structure, recent observations with modern instruments with an excellent time, frequency, and angular resolution have revealed many different FS in shock radio bursts \citep{Magdaleni_2020}. These FS are said to be a signature of turbulence that shocks encounter while propagating through the inhomogeneous turbulent corona \citep{2018_chen,Carley_2021}.
The propagation effects modify the intrinsic properties (shape,  size, positions, brightness temperature) of the observed radio source \citep{2017kontar,Kontar_2019,Sharma_2020, 2021Zhang,Ryan_2021,2021Pearse}. 
There have been several studies \citep{1965BFokker,1971Steinberg, 1972Riddle,2021Zhang} which model the extent of the scattering effects on the radio waves in the corona and some which compare the simulations with the observations of solar radio bursts \citep[e.g.][]{Kontar_2019}.
There is currently an ongoing debate on the length scales of turbulence that cause radio wave scattering. 
If the scattering happens where the spectrum is purely Kolmogorov-like, then the characteristic scale length of the density fluctuations is considered to be a combination of the inner and outer scale of the turbulence spectrum \citep{2007Thejappa}. Alternatively, it is also thought that the radio scattering happens near the inner scale.  In the literature, there is no consensus on which length-scale to use \citep{1994Bastian}. The lack of consensus is partly due to the lack of direct diagnostics of this effective length scale.



In this paper, we analyze a variety of fine-scale structures and fine-scale motions of type II radio bursts using the Murchison Widefield Array \citep[MWA;][]{Tingay2013}. Firstly, we provide rare imaging of band-splitting that does not agree with the upstream/downstream scenario, supporting the \cite{mclean1967} model.
Secondly, we developed a new technique that uses MWA imaging spectroscopy to examine fine-scale source motions of the type II burst and examine this in the context of coronal turbulence. Section 2 provides an observational overview of the type II burst, Section 3 describes the methods of the data analysis. The results and discussions are presented in Sections 4 and 5, respectively, followed by conclusions in Section 6.




\section{Observations}

   \begin{figure*}
   \centering
      \includegraphics[width = 16cm]{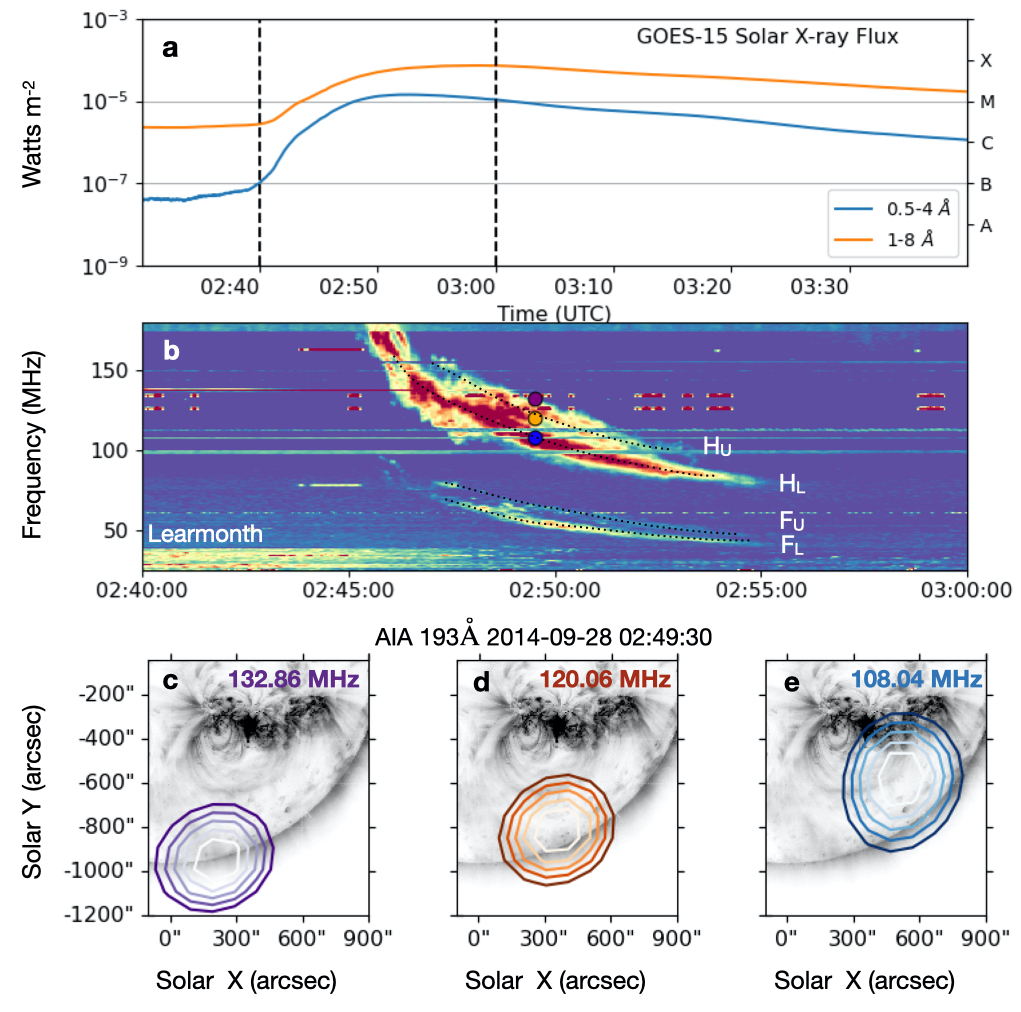}
      \caption{a) a) GOES-15 light curves indicating a M5.1 class flare. The two vertical dashed lines show the time range of the Learmonth dynamic spectra.
      b) The dynamic spectra of the type II radio burst observed by the Learmonth spectrograph. It is clear that the fundamental band show band-spitting where the high and low frequency branches are indicated by F\textsubscript{U} and F\textsubscript{L}, respectively. The corresponding band-splitting in harmonic band is marked by H\textsubscript{U} and H\textsubscript{L}, respectively.
      (c,d,e) The three images show the position of MWA radio contours overlaid on the AIA 193  \AA  ~ channel base image at 02:49:30\,UT. The purple, orange, and blue contours are the MWA 132.86, 120.06, and 108.54 MHz contours (50\%, 60\%, 70\%, 80\%, 90\% of the peak intensity), respectively. They show the position of the radio sources of different frequencies at the same time, indicated by purple, orange, and blue points on the Learmonth spectra.} 
         \label{fig:goes_lea}
   \end{figure*}

On 28/09/2014, a Geostationary Operational Environmental Satellite-15 (GOES-15)\footnote{\url{https://space.oscar.wmo.int/satellites/view/goes_15}} M5.1 class solar flare started at 02:39:00\,UT from the active region (AR) NOAA 12173 (Fig~\ref{fig:goes_lea}a).
This flare was also associated with a slow CME which was first seen by the Large Angle and Spectrometric Coronagraph C2 \citep[LASCO;][]{LASCO1995} at 03:24:05\,UT. According to the LASCO C2 CDAW\footnote{\url{https://cdaw.gsfc.nasa.gov/CME_list/nrl_mpg/2014_09/140928_c2.mpg}} catalogue, the linear speed of the CME was 215\,km\,s$^{-1}$ on the plane-of-sky. This eruption event was observed in extreme ultraviolet (EUV) wavelengths by the Atmospheric Imaging Assembly  \citep[AIA;][]{2012Lemen} onboard the Solar Dynamics Observatory \citep[SDO;][]{2012Pesnell}. 

At 02:45\,UT, the peak time of the X-ray flare, Learmonth Spectrograph\footnote{Learmonth Observatory; \url{https://www.sws.bom.gov.au/Solar/3/1}} observed a type II burst, the radio signature of a shock during this eruptive event (Fig~\ref{fig:goes_lea}b). The dynamic spectra shows both fundamental (f$_{p}$) and harmonic ($\sim$2f$_{p}$) of the burst. 
Type II radio bursts often show band-splitting \citep[e.g.][]{vrsnak2001}, where each of the fundamental and harmonic emissions are split in to two sub-band which show the same spectral drifts. We refer to these sub-bands as the high frequency sub-band (HFS) and low frequency sub-band (LFS).
Visually it can be difficult to characterize the HFS and LFS in the type II harmonic band. To make the band-splitting clearer the Learmonth dynamic spectra had been smoothed using a small kernel (2x2 pixels) in Fig~\ref{fig:goes_lea}b. The fundamental band shows the clear band-splitting where the HFS and LFS are indicated by $F_{U}$  and $F_{L}$. We see the corresponding band-splitting in the harmonic band, indicated by $H_{U}$  and $H_{L}$ . 
The drift rates computed for both branches in the harmonic band are found to be very similar. The drift rates of the HFS and LFS are -0.160 and -0.158 MHz/s, respectively. Hence we regard the two observed sub-bands to correspond to the band-splitting, often seen in type II radio bursts.

This type II burst was observed using the
MWA, a low-frequency radio interferometer capable of providing a time and spectral resolution of 0.5\,s, and 40\,kHz respectively, across its 80-300\,MHz observing band.
The MWA observed the harmonic band of this radio burst in 6 different spectral pickets of 2.56\,MHz bandwidth each.
These observations spanned the frequency range from 79.10 to 133.36 MHz with a spectral resolution of 40 kHz. These spectral bands of MWA observations are marked with black rectangular boxes on the Learmonth dynamic spectra in Fig~\ref{fig:goes_lea}b.

\section{Methods}\label{methods}
The high fidelity solar radio images from the MWA data were obtained using the Automated Imaging Routine for Compact Arrays for the Radio Sun \citep[AIRCARS;][]{Mondal_2019}.
 AIRCARS is a self-calibration based radio imaging pipeline optimised for spectroscopic snapshot imaging for centrally condensed arrays like the MWA. Its efficacy has been demonstrated in multiple recent works requiring high imaging quality and applications ranging from the first ever detection of mSFU level metrewave impulsive emissions from the quiet Sun \citep{Mondal2020b} to detection of spatially resolved gyrosynchrotron emission from CMEs \citep{Mondal2020a}, multiple investigations of quasi-periodic-pulsations in active solar radio emissions \citep{Mohan2019b, Mohan2019a, mohan2021a, mohan2021b, mondal2021a} and for developing precise absolute solar flux density calibration techniques \citep{DK-2022}.
Imaging and calibration was done using AIRCARS at a time resolution of 0.5\,s and spectral resolution of 40 kHz for 8 spectral channels in each of the 6 MWA spectral sub-bands. We have applied independent gains obtained from self-calibration to each spectral channel.

Flux calibration was done using the fortuitous presence of Virgo-A in the very large field of view of the MWA.
The flux density of Virgo-A at our frequencies of observation was obtained using a linear spectral fit to the data available from the NASA/IPAC Extragalatic Database\footnote{NED; https://ned.ipac.caltech.edu} and a model for the MWA beam available from \citet{Sokolowski2017}.\par
Figure~\ref{fig:goes_lea}c-e shows some examples MWA radio images of the type II emission at 132.86 (purple), 120.06 (orange) and 108.54 (blue) MHz.
The radio contours are overlaid on the AIA 193\,\AA  ~base image of the eruption event at 02:49:30~UT. 
The frequencies and time of these radio sources are also marked on the Learmonth spectra by purple, orange, and blue circles. The radio contours show that the type II harmonic emissions have a single compact source which can be well described by 2D elliptical Gaussian. Hence, a Gaussian model was fit to each of the AIRCARS images for all 6 spectral bands by doing chi-square minimization using the Levenberg–Marquardt algorithm \citep{levenberg,Marquardt} and the parameters of the best fit Gaussian models used to extract the information about the location, size, shape, and intensity of the burst emission. The fitting shows the emission source has
semi-major axis in the range 6\arcmin--10\arcmin and semi-minor axis lying between 5\arcmin--7\arcmin. 
The uncertainty in the position of the fit was calculated using CASA imfit software which uses the standard method described in \cite{Condon_1997}. According to this method, the uncertainty in position depends on full width at half maximum (FWHM) of major and minor axes, peak intensity, the RMS noise, and the pixel width of the direction coordinate. The uncertainties in position are found to be very small, less than 1.2\arcsec, due to the high signal to noise ratio of the data.

 The positional information from the Gaussian fits is used in the following sections to understand the kinematics of the split band sources and the fine scale motion of type II.

\begin{figure*}
   \centering
      \includegraphics[width = 17cm]{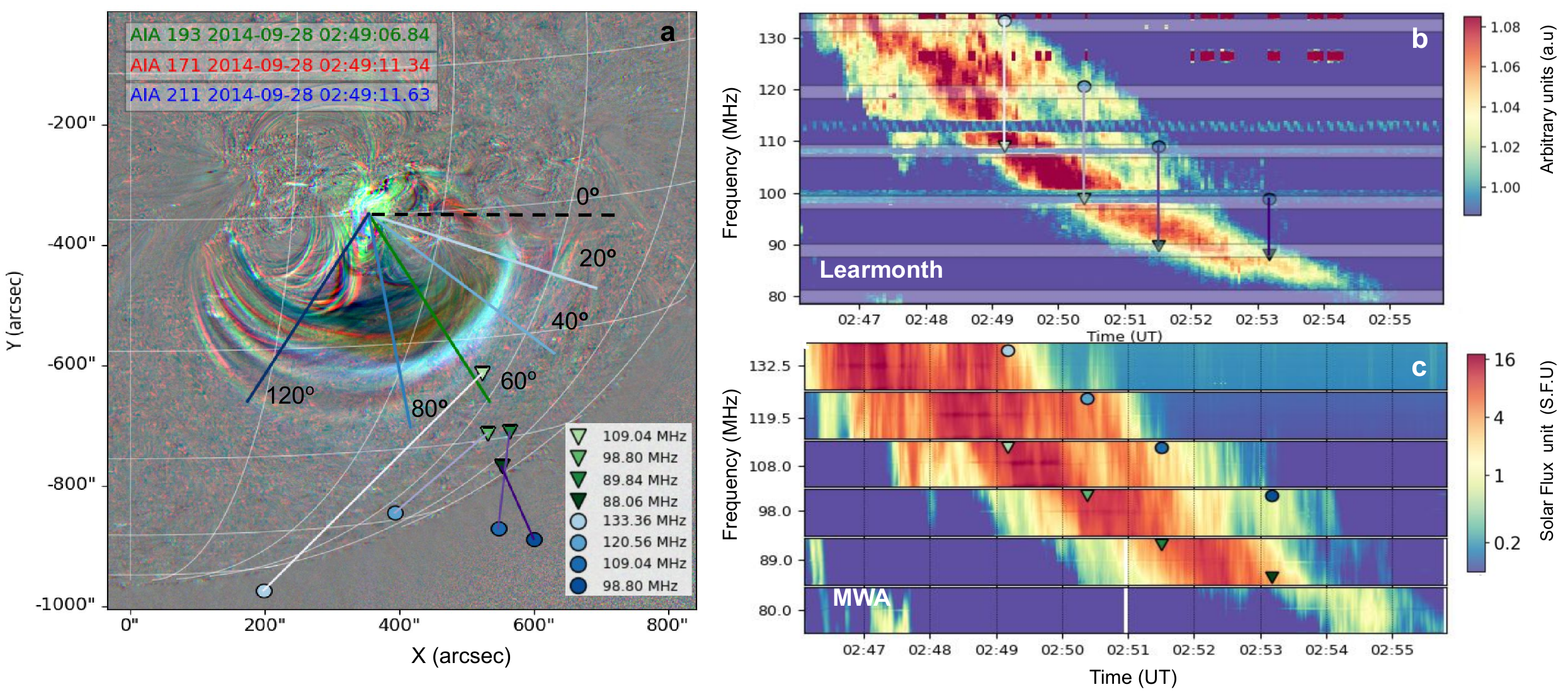}
      \caption{a) The ratio image of the eruptive event with SDO/AIA in 171, 193, 211  \AA  ~channels ~at 02:49:11 UT. The blue circle and green triangle points on the image are the position of radio sources of different frequencies at the same time from HFS and LFS, respectively (shown in panel b). The lines connect the HFS and LFS points at the same time. It is clear that both band sources are moving in different directions from each other and  they are quite separated in the position.
      The top (b) and bottom (c) figure in the right panel show the Learmonth and MWA dynamic spectra of the type II harmonic band. The rectangular panels on the Learmonth spectra highlight the regions of the type II MWA has observed.} 
         \label{fig:band-split}
   \end{figure*}

\section{Results}
\subsection{Kinematics of the split-band radio sources}
Figure~\ref{fig:band-split}b-c shows the harmonic band of the type II burst in the Learmonth and MWA spectra. Points of different frequencies at the same time have been chosen from HFS (blue circles) and LFS (green triangle) of the harmonic band in the dynamic spectra. Figure~\ref{fig:band-split}a shows radio source positions (from the 2D Gaussian fits) of these frequencies and times overplotted on the AIA image, a three colour running ratio image of the eruption produced by AIA 171, 193, 211  \AA  ~channels ~at 02:49:11\,UT.
Each AIA image was first normalised by its exposure time and the time difference in images for the ratio was $\sim$3 mins.
The observations show that the LFS and HFS are spatially separated by up to around 485\arcsec  ($\sim$372\,Mm), and get progressively closer at 157\arcsec ($\sim$120\,Mm) on the plane-of-sky (POS). The uncertainty in the position of the sources was considered to be the average position fluctuations of the radio sources $\sim$30\arcsec (see Fig~\ref{fig:histogram}).
The separation of split-band sources occurring simultaneously is shown by the points connected by purple lines in Fig~\ref{fig:band-split}a.
The split-band sources also have noticeably different kinematics, with LFS (green triangles) propagating towards south west, the most prominent direction of the eruption, and HFS (blue circles) propagating in a lateral, east-west direction. Towards the end of the type II, the HFS is positioned at a greater radial distance (in the plane-of-sky). 
The separation in space of the LFS and HFS by $\sim$372\,Mm is clearly indicative of the two radio sources being generated in different regions of the corona. This observation clearly contradicts the upstream/downstream scenario e.g. the HFS and LFS sources should be closely positioned in space and move in the same direction.
In Section \ref{sec:discussion}, we further discuss these source kinematics with regard to the various hypotheses for the band-splitting phenomena and explain that the source positions and separations that we observe do not support the upstream-downstream hypothesis for this event.  
 
\subsection{Large-scale kinematics of the shock}
\begin{figure}[!b]
   \centering
      \includegraphics[scale = 0.55]{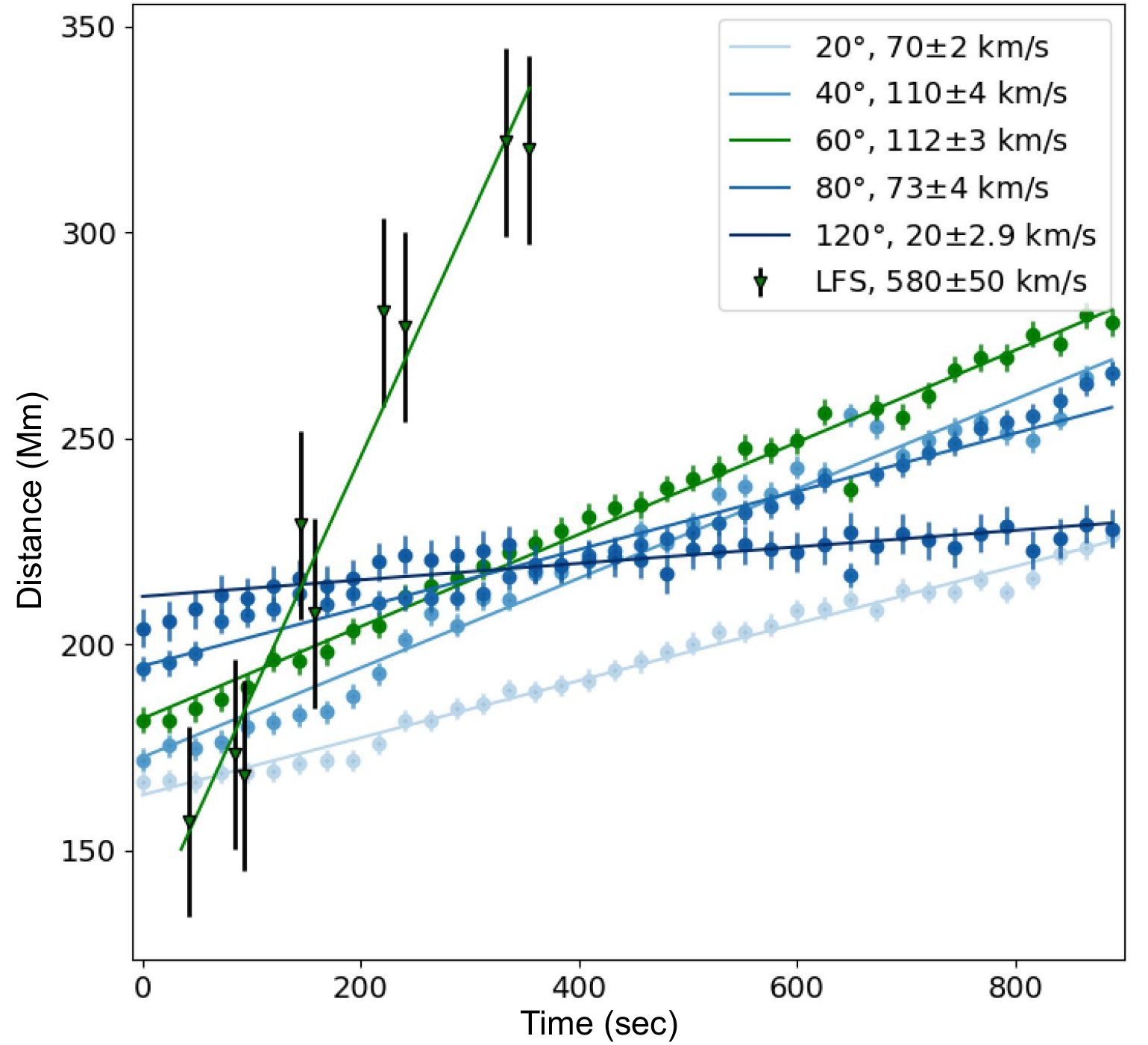}
      \caption{The kinematics of the radio sources with respect to eruption. Given that the speed of the shock is much higher than the maximum speed of the eruption event, the shock originating this burst is expected to be a piston driven shock where the shock speed can exceed the driver speed. }
         \label{fig:kinematic}
   \end{figure}

As the next step, in Fig~\ref{fig:kinematic}, the speeds of the radio sources and the eruption event in EUV observations were calculated to understand the relationship between the eruptive active region and the type II burst.

Since the propagation of LFS sources was closely following the direction of the eruption event at around 60 degrees (green line on the AIA ratio image in the Fig~\ref{fig:band-split}), these sources have been chosen to calculate the speed of the radio sources. 

The speed of the eruption in EUV was calculated in several directions indicated by lines on Fig~\ref{fig:band-split}a at 20, 40, 60, 80, and 120 degrees from the black dotted line (in the plane of the sky). 
The brightest point on the EUV front for these directions is chosen for each time using point-and-click method. The distance of these points was calculated from the location of the active region origin and linear fits are performed over these distances using linear least-squares regression, shown in Fig~\ref{fig:kinematic}. It was found that the speed of the EUV front is not the same in every direction and varies from 20 - 112\,km\,s$^{-1}$. The uncertainty in the distance of these points is calculated from point-and-click uncertainty over several attempts. 
The eruption was most prominent towards 60 degrees and, where its speed was 112$\pm$3\,km\,s$^{-1}$. 

In Fig~\ref{fig:kinematic} the green triangles indicate the position of the radio sources over time. For the position uncertainties for each radio source, the average fluctuations of the sources has been taken into account which is $\sim$0.5\,arcmin or 23\,Mm (Fig~\ref{fig:histogram}). To determine the speed, the distance of the radio sources has been calculated from the active region position and the linear regression has been performed to get the speed of the radio sources and the error in the speed. The type II radio source speed (shock speed) is found to be 580$\pm$50\,km\,s$^{-1}$, which is much faster than the maximum driver speed of 120\,km\,s$^{-1}$ observed in EUV\footnote{Note the plane-of-sky CME linear speed  was 215\,km\,s$^{-1}$, reported by SOHO/LASCO CDAW catalogue. This speed deprojected by the active region angle from the sky-plane is $\sim$ 430\,km\,s$^{-1}$.}.

Given the much larger shock speed than driver speed, we expect that the shock causing this type II burst is the piston driven mechanism \citep{2008Vrsnak,2011Nindos,Magdaleni_2020}. This is discussed further in Section~\ref{discuss1}.
   
\subsection{Fine-scale motion of radio sources}

\begin{figure*}
   \centering
      \includegraphics[width = 17 cm]{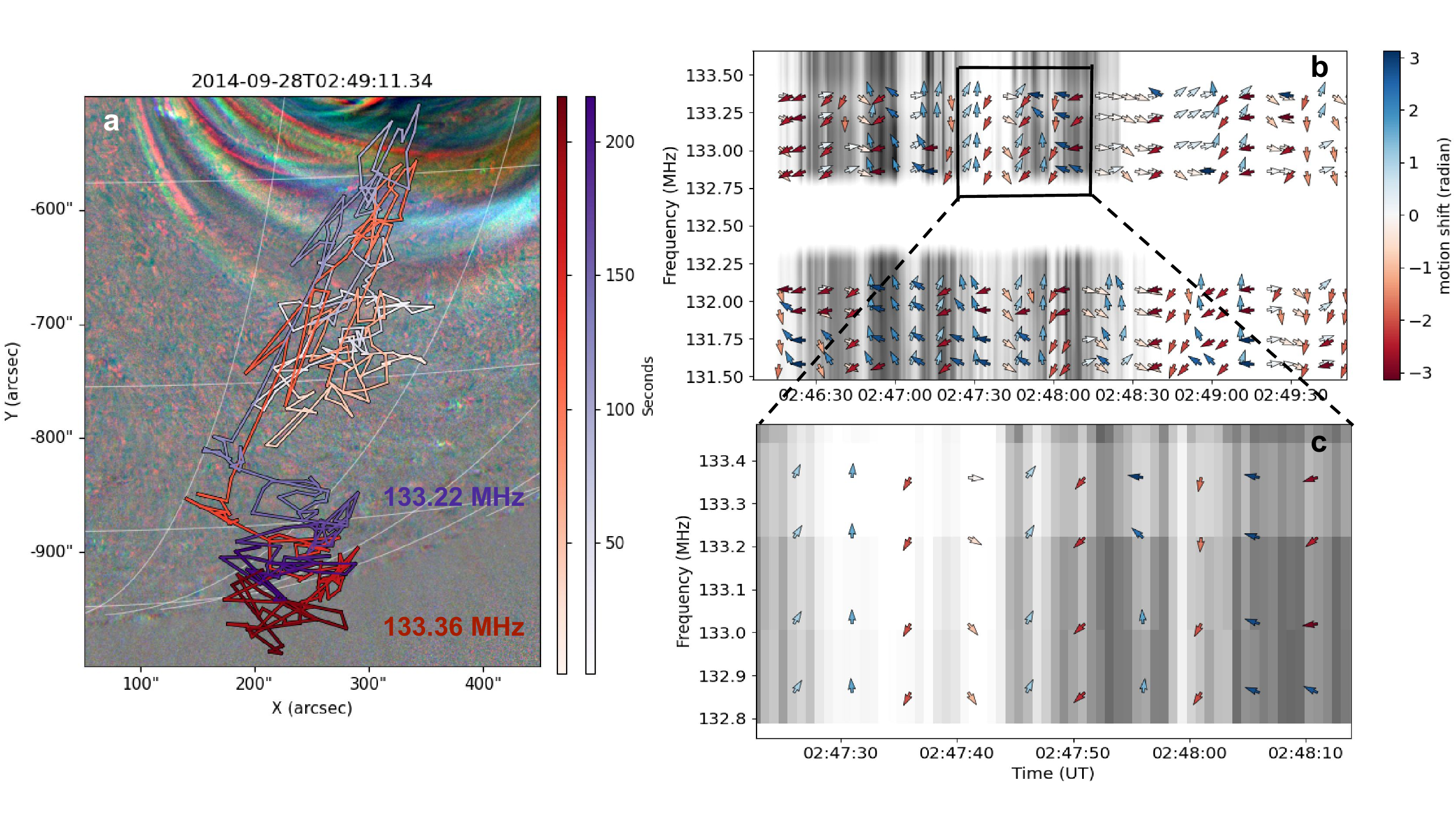}
      \caption{a) The running ratio image of the eruption in AIA 193, 211, and 171  \AA  ~channels ~at 02:49:11 UT.
      The position of the type II sources of 133.36 MHz (red) and 133.22 MHz (purple) have been plotted over the image for a duration of ~5 min, represented by both red and purple color bars.  The dynamic spectra in panels (b) and (c) on the right show the peak amplitude of the Gaussian fitted source. The arrows represent the instantaneous change in direction ($\theta_{i+1}-\theta_{i}$) of the source motion for different frequencies (131.58, 131.74, 131.94, 132.08, 132.86, 133.02, 133.22 and 133.36 MHz) in the channel. The blue arrows represent when the change in angle lies between 0 to $\pi$ radian. Similarly, the red arrows represent when the change in direction lies between 0 to -$\pi$ radian. Panel (c) shows a zoomed-in view of the region marked in panel (b). In panel (c), one can see at each time the arrows point in almost the same direction across nearby frequency channels, exhibiting the correlated fine scale motion.}
         \label{fig:arrow}
   \end{figure*}
\begin{figure*}
   \centering
   
      \includegraphics[scale=0.3]{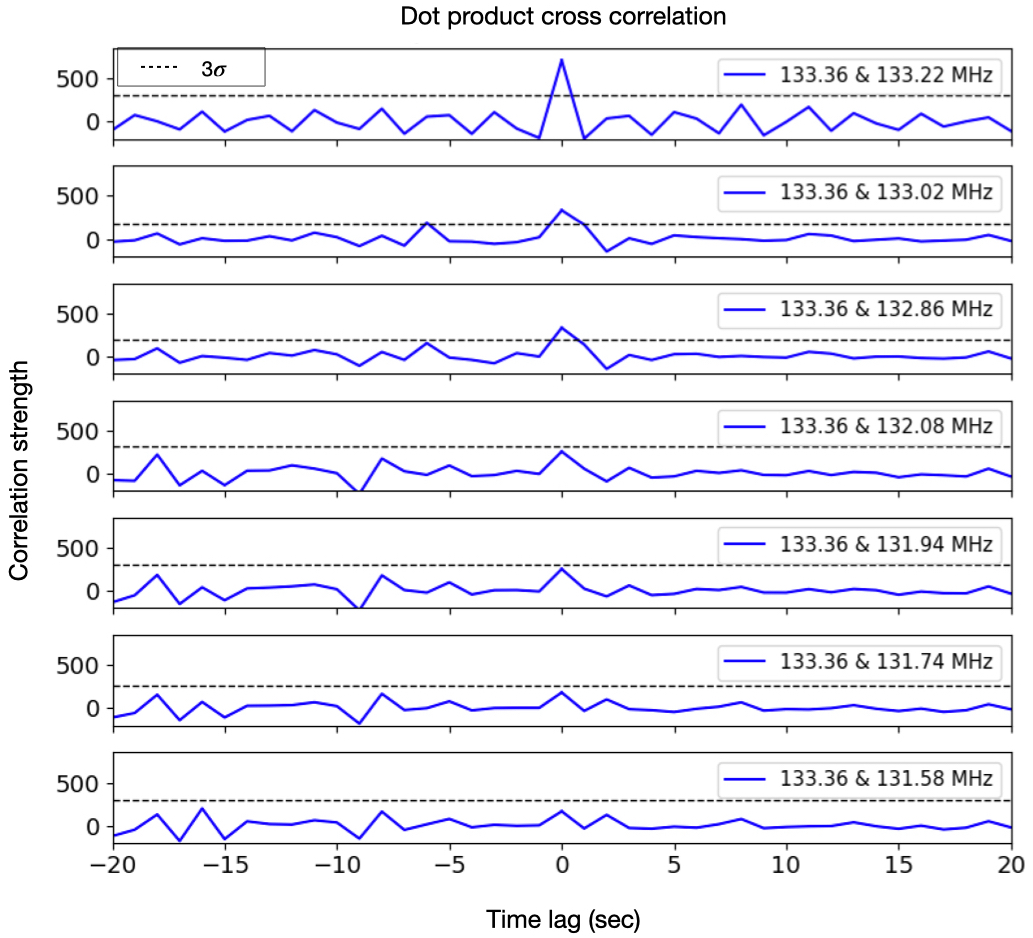}
      \caption[Cross correlations]{Dot product correlation (blue) of the magnitude and $\Delta\theta$ between 133.36\,MHz and seven other frequencies (133.22, 133.02, 132.86, 132.08, 131.94, 131.74 and 131.58\,MHz). Dot product correlation represents a cross correlation of instantaneous direction and amplitude changes for sources in adjacent frequency channels. The dashed line represents the $3\sigma$ above the background noise for each of the correlation plot. We regard the correlation to be significant only if it exceeds $3\sigma$. Clearly, there exists a significant correlation between the top three frequency pairs at zero time lag, which drops below $3\sigma$ level with increasing separation between frequencies}
      \label{fig:correlation}
   \end{figure*}

Figure~\ref{fig:arrow}(a) shows the harmonic type II source motion at 133.22 (purple) and 133.36 (red) MHz with respect to the active region for the duration of the burst, starting at 02:46:12 UT and ending at 02:49:50 UT. The background image is the three colour ratio image of the eruption in AIA 193, 211, and 171  \AA  ~channels ~at 02:49:11 UT. Clearly, the fine scale motion of the radio sources shows a complex chaotic character. We describe this motion by the parameter $\Delta\theta$, the instantaneous change in source direction at time t+1 given by $\Delta\theta_{t+1} = \theta_{t+1}-\theta_{t}$. 
This direction change is expressed in radians and shown for each frequency channel (131.58, 131.74, 131.94, 132.08, 132.86, 133.02, 133.22, and 133.36 MHz) by the arrows on the dynamic spectra in  Fig~\ref{fig:arrow}b-c. 
The background dynamic spectra (grey color scale) in both figures represent the peak amplitude of the Gaussian fitted source at the corresponding frequency.
Figure ~\ref{fig:arrow} (c) shows a zoomed-in view of the region marked by the black rectangle on the Fig ~\ref{fig:arrow}(b). While  the arrows indicate random instantaneous position changes at each time step, we observed that the position changes behave similarly for adjacent frequencies. This implies that the fine-scale motion is correlated across nearby frequencies. To investigate this systematic correlated behavior, we produce a cross correlation of instantaneous direction and amplitude changes for sources in adjacent frequency channels. For example, $\vec{f}(t)$ represents the direction and magnitude of the instantaneous change in position over time for one frequency, while $\vec{g}(t)$ represents the same for another frequency. We use a dot-product correlation to look for correlated position change (magnitude and direction) between nearby frequencies, given by 
\begin{equation}
\vec{f}*\vec{g}(\tau)=\int \vec{f}(t) \cdot  \vec{g}(t+\tau)dt = \int f(t)  g(t+\tau)\cos(\phi(t))dt.
\end{equation}
where $\phi(t)$ is the change of direction between $\vec{f}(t)$ and $\vec{g}(t)$.
This equation will show a large response when the position shift direction and magnitude of the two time series for two different frequencies are the same.
Figure~\ref{fig:correlation} shows the dot-product correlation between 133.36\,MHz and seven other frequencies (133.22, 133.02, 132.86, 132.08, 131.94, 131.74 and 131.58\,MHz). The dashed line represents $3\sigma$ above the background noise for each of the correlation plot. We consider there is a significant correlation only when the correlation peaks are above $3\sigma$.
We can see a significant correlation at zero time lag between 133.36 and 133.22\,MHz. The correlation steadily drops with increasing frequency difference from 133.36\,MHz and drops below significance beyond the nearest three frequencies. 

\begin{figure}[!b]\label{cor_all_chan}
   \centering
   
      \includegraphics[width=8cm]{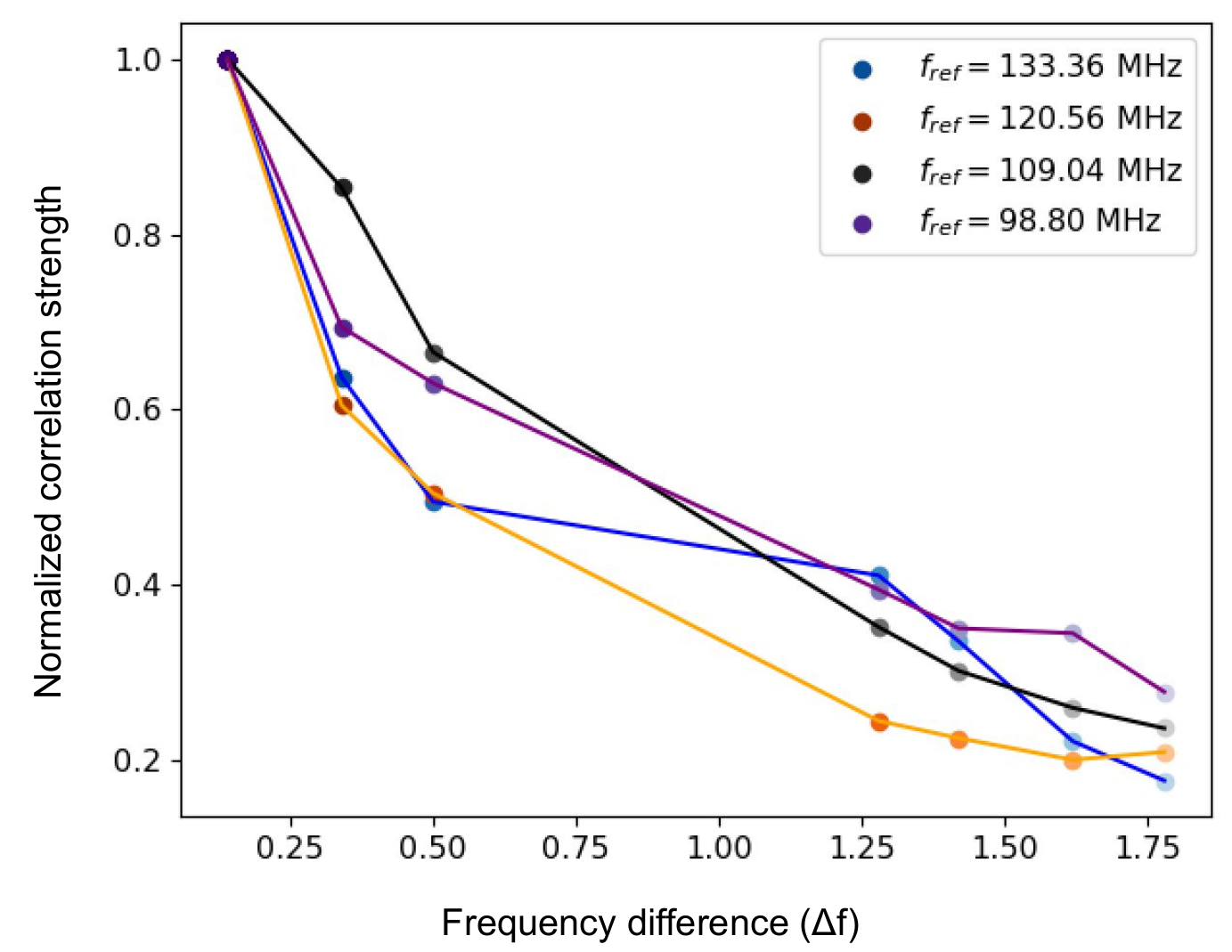}
      \caption[Cross correlations for all channels]{Normalized correlation strength vs. frequency difference for all 4 coarse channels. 
      Here the frequency difference ($\Delta$f) is the difference between the reference frequency (f$_{ref}$) and each of the other frequencies for the dot product cross correlation in MHz.
      For each channel, f$_{ref}$ is chosen to be the highest frequency and each circle points represent the correlation strength for each frequency in that channel.
      The correlation strength decreases with increasing $\Delta$f.
      }
      \label{fig:corr_all_chan}
   \end{figure}

In Fig~\ref{fig:corr_all_chan}, we show the correlation strength as a function of frequency difference ($\Delta f$) for four coarse channels. The frequency difference ($\Delta f$) is the difference between the reference frequency  (f$_{ref}$; chosen to be the highest frequency in the channel) and each frequency in the same channel used for the correlation. Each series of points (indicated by different color) represents the correlation between f$_{ref}$ and f$_{ref} + \Delta f$ frequency for each different coarse channel. The color scale of the circles for each coarse channel represents the $\Delta f$ for eight frequencies in each channel. Clearly, the correlation strength decreases for larger $\Delta f$ from f$_{ref}$ for each coarse channel.


The phase calibration solutions are independent across frequency channels, so the fine-scale motion among different frequency sources would naively be assumed to be uncorrelated. 
However a correlation is found to exist in fine scale motion among nearby frequency channels and this correlation must have a physical cause in the corona.
Section~\ref{fine} discusses this physical cause and its potential relation to radio source scattering in the corona due to turbulence.
   
\begin{figure*}
   \centering
      \includegraphics[scale=0.35]{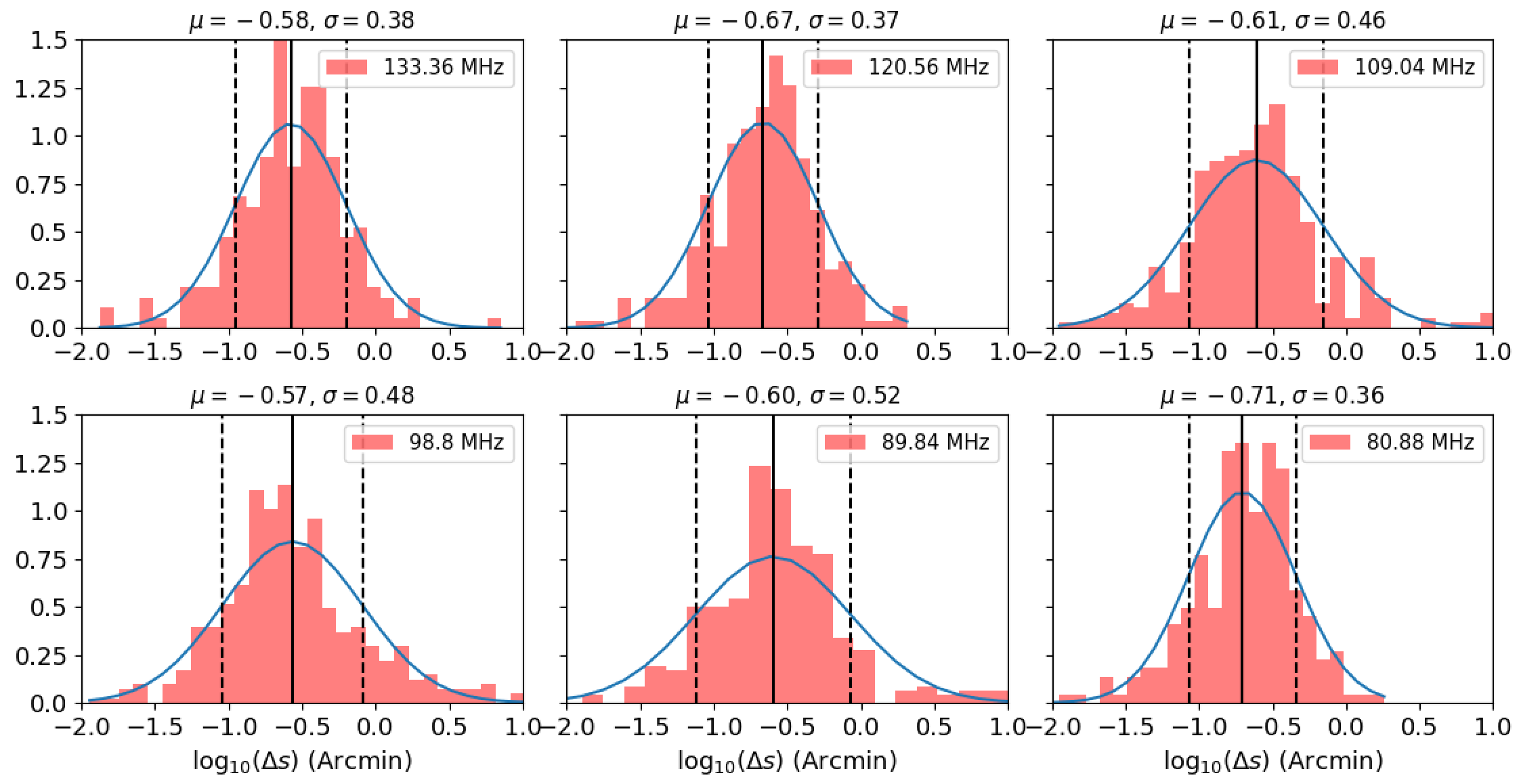}
      \caption{Histogram plot showing the probability density of instantaneous magnitude change in position ($\Delta s_ {t+1}= s_{t+1}-s_{t}$) of sources for a single frequency from all six channels. The blue curved line in each plot represents the Gaussian fitting done on the histogram. The mean and 1 sigma of the Gaussian are indicated as solid and dotted black line, respectively.}
         \label{fig:histogram}
   \end{figure*}

Figure~\ref{fig:histogram} shows histograms of the instantaneous position shift magnitude ($\Delta s_{t+1} = s_{t+1}-s_{t}$) of sources for a single frequency from all six coarse channels. The $x$-axis of all the plots is in log-space and a Gaussian has been fitted over the histogram for all channels to calculate the mean (black solid line) and standard deviation (black dotted line) of $\Delta s$.
The distributions show a clear log-normal distribution in each coarse channel, and 
provide a clue to the potential physical cause of the stochastic motion observed in images as discussed in Sec. \ref{fine}. 


\section{Discussion} 
\label{sec:discussion}
\subsection{What causes band-splitting in this type II event?} \label{discuss1}
In this work, we analyse a split-band metric type II burst that spatially and temporally relates to the 28th September 2014 eruption event, a M5.5 solar flare followed by a slow CME. The EUV observations of this event show the propagation of the expanding loops of the active region (Fig~\ref{fig:band-split}, left panel). It is found that the disturbance caused by the expansion is not propagating uniformly in all directions and the maximum speed progressed towards the south-west at $\sim$112\,km\,s$^{-1}$. However, it should be noted that there is no direct evidence of the shock wave in the AIA observations. The direct evidence of this shock front, a split band type II burst, was found in the dynamic spectra from the Learmonth observatory and the MWA.


The LFS sources are found to propagate towards the same direction as the eruption and the speed of the radio sources is found to be around 580\,km\,s$^{-1}$. This is considerably faster than the driver speed of $\sim$112\,km\,s$^{-1}$ derived from the EUV observations. It is therefore likely that the shock associated with the eruption was a piston driven shock \citep{2008Vrsnak}. In such a scenario, a subsonic driver can cause a shock. The plasma is pushed outwards by a 3D piston impulsively, and the plasma ahead of the piston (expanding magnetic loops) cannot pass behind the driver. As a result, the shock `races away' from the driver, and shock speed is greater than the driver speed. This is in contrast to the bow shock scenario, in which plasma can pass behind the driver, and the shock and driver speed are the same.

High fidelity interferometric imaging of MWA provides a golden opportunity to image the source location of the split bands of type II at the same time. There have only been a few studies that achieved this \citep{zimovets2012,Zucca_2014,2015Zimovets,Chrysaphi2018}. This allows us to understand the cause of band-splitting and also to compare the two widely accepted band-splitting theories. The most popular interpretation due to \citet{smerd1974,smerd1975} is that the high and low frequency split bands are coming from downstream (behind) and upstream (ahead) regions of the shock front at the same time, respectively. The density compression across the shock (see Section 2.1.2) leads to a separation of plasma frequency of upstream-downstream sources ($f_p\sim 8980 \sqrt{n_e}$). There has been some evidence supporting this interpretation \citep{Chrysaphi2018,zimovets2012,Zucca_2014}. There have also been results that are not in the favor of this upstream/downstream scenario. \cite{Du_2014} found that the spectral features first appeared in the HFS seconds earlier than in the LFS which, according to the upstream/downstream scenario, should happen the other way around. According to the upstream/downstream scenario, there should also be a strong correlation between the shock velocity and the frequency ratio between the sub-bands, but \cite{Du_2015} show only a very weak or no correlation between these two parameters. The review work by \cite{2011Cairns} also argues that in the downstream region the absence of energetic electrons with respect to the ambient plasma does not lead to the growth of Langmuir waves in the region, not supporting the upstream/downstream scenario. 
An alternative mechanism was proposed by \cite{mclean1967} where the band-splitting originates due to multiple parts of the shock front encountering coronal environments of different physical properties, such as variations in electron density or magnetic field. There have been very few observational studies that support this view \citep{2015Zimovets}.

In our analysis, it is observed that the apparent location of the split band sources are initially rather far apart ($\sim$372\,Mm in POS) then get gradually closer with time ($\sim$120\,Mm in POS). 
They also appear to move in different directions with time.
For the reasons outlined below, the large observed spatial separation and their motion cannot be attributed to propagation effects alone.
This emission is at the harmonic (f = 2f$_{p}$), where the impact of the propagation effects is much less dramatic as compared to the fundamental.
For instance, \cite{2021Zhang} show that depending on the anisotropy and density fluctuations, a harmonic source at 35\,MHz can suffer absolute maximum position offset of $\sim$ 0.1 $R_{\odot}$ ($\sim$69\,Mm). The observed frequency range of the type II burst studied here higher (79-134\,MHz), implying that the propagation effects should be smaller still. 
Even when the propagation effects lead to large changes in the apparent locations and sizes of the observed sources \citep[e.g.][]{Sharma_2020}, these changes are expected to be similar at nearby frequencies.
The difference in observing frequencies for the HFS and LFS is not large. It lies in the range of 10 -- 20 MHz leading to a $\Delta f/f \sim 15\%$.
Despite their small spectral separation, the observed differences in locations and motions of the HFS and LFS sources are large. This implies that these differences can not arise due to propagation effects alone and must come from these sources being intrinsically well separated.

\begin{figure}[!b]
   \centering
      \includegraphics[scale = 0.25, trim = 2cm 0cm 0cm 0cm]{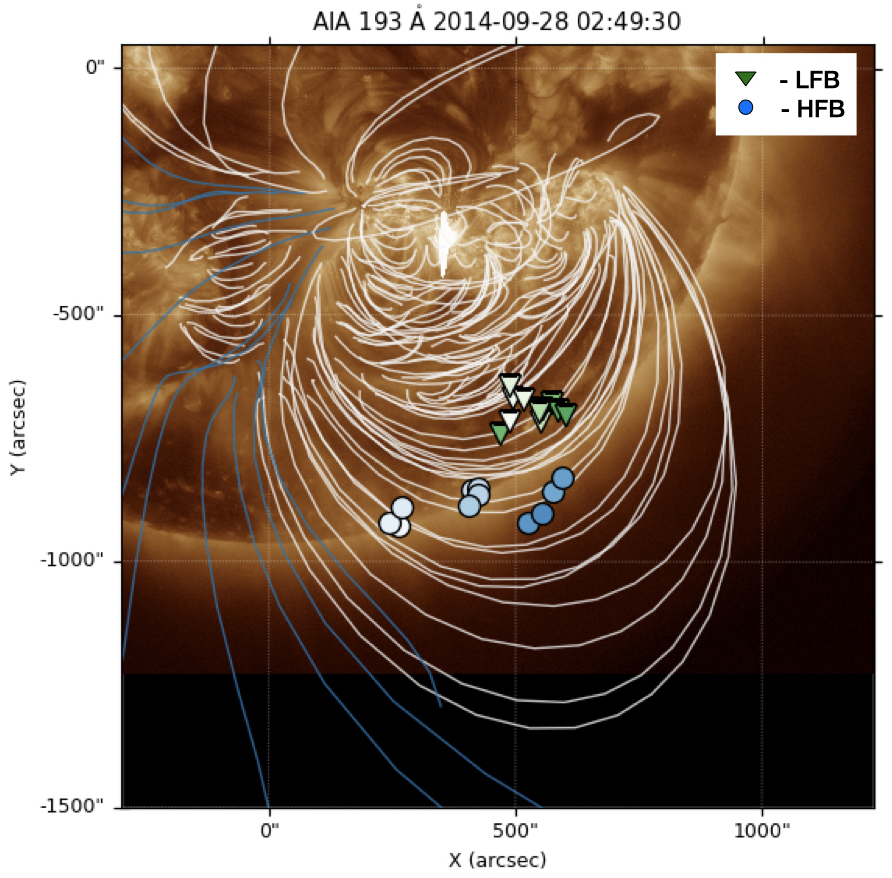}
      \caption{Magnetic field lines extrapolated with PFSS model overlaid on the region associated with the eruptive event on the AIA 193 \AA ~image at 02:49:30 UT. In our region of interest the field lines mostly are closed magnetic field loops (white). Clearly, LFS and HFS sources are associated with different loops. }
         \label{fig:pfss}
   \end{figure}

To understand the differences in these regions in terms of the geometry of the magnetic field lines, the extrapolation of the field lines is done with the Potential Field Surface extrapolation (PFSS) model \citep{pfss_2019} using the Global Oscillation Network Group\footnote{\url{https://gong2.nso.edu/archive/patch.pl?menutype=s}} (GONG) observations at 02:14\,UT. The PFSS field lines are overlaid on the AIA image of the eruption event at 02:49:30\,UT in 193 \AA ~channel in the Fig~\ref{fig:pfss}. Both sub-band sources of the type II are on different magnetic field loops, with the LFS closer to the active region core than the HFS. As discussed earlier there can be a significant difference in the true and apparent position of sources. It is however clear from Fig~\ref{fig:pfss} that the separations are so large that these sources lie in different coronal environments.

This implies that the radio emissions must come from different parts of the shock-front causing two sub-bands in the type II dynamic spectrum (Fig~\ref{fig:band-split}b). The analysis by \cite{Zucca2018, 2021Kouloumvakos} also suggests a similar conclusion.

This is rare evidence against the commonly used upstream-downstream interpretation of split-band sources used widely to estimate the shock Mach number \citep{vrsnak2001} and the strength of the magnetic field \citep{2017Kumari,2019Kumari}. Our work cautions against using such a diagnostic for split-band sources and reiterates the need for full imaging-spectroscopy in interpreting type II fine structure. 
   
\subsection{Fine-scale motions: what is the case of the perturbations?} \label{fine}

Figure~\ref{fig:arrow} shows the complex motion of the type II sources over time. 
Our analysis showed this motion, although disordered and random, exhibits a systematic correlation across nearby frequencies. Figure~\ref{fig:correlation} shows that the source motion is well 
correlated across a bandwidth of 0.5\,MHz for a specific spectral channel and this is true for all other channels. 
As discussed later in this section, the correlations in motion cannot arise due to calibration effects, as independent calibration solutions are applied to each frequency separately. This then poses the question of what causes disordered, instantaneous position shifts to be correlated across radio sources at nearby frequencies?


Recent studies have shown that low frequency radio sources can appear to be displaced from their true location due to propagation effects arising from the density turbulence in the corona \citep[e.g.][]{Kontar_2019,2021Zhang}. 
Furthermore, previous studies  have shown that stochastic fluctuations in turbulent velocity fields can show a log-normal distribution \citep[e.g.][]{arneodo_1998,Mouri_2009}. 
For each frequency channel we observe here, the distribution of instantaneous magnitude change in position ($\Delta s_ {t+1}$) exhibits such a log-normal distribution (Fig~\ref{fig:histogram}). 
It is possible that the instantaneous and stochastic source displacements we observe here may arise due to turbulent density fluctuations. 
For example, radiation at nearby frequencies launched from the same neighbourhood is expected to encounter essentially the same turbulent structures in the corona and consequently follow similar ray paths.
Hence, we interpret this observed correlation over small spectral spans as arising due to propagation effects like scattering and refraction.
An additional possibility is that these apparent motions are arising due to small scale local variations of the shock surface itself. 
Simulations have shown that a turbulent corona can cause a shock surface to become rippled and inhomogenous, with shock properties such as shock strength and acceleration efficiency being inhomogenous \citep[e.g.][]{Andreopoulos_2000,Zank_2002,lowe2003,neugebauer2005,Burgess_2006,Giacalone_2008,Lu_2009,2011Vandas,Guo_2010}.
If this is indeed the case, the root cause for the observed rapid and small scale changes in the position of the emission is still the turbulence in the background corona. 
This implies the observed spectral separation over which there is correlated motion carries information of the length scale of the coronal turbulence affecting the shock surface. 


We relate the spectral separation of correlated fine channels to a spatial separation using a coronal electron density model. We have chosen the Saito density model \citep{1970Saito}.
A correlation among the fine channels separated by 0.5\,MHz represents a correlation among radio sources separated in space by 1 - 2\,Mm.
We note that though the absolute coronal heights might differ by larger amounts between different coronal electron density models, the differences between the coronal heights corresponding to two nearby frequencies are much smaller. For example, we followed the same procedure with the Newkirk density model \citep{1961Newkirk} and the resulting spatial separation is the same. So these numbers hold independent of the electron density model used.
We interpret this as the correlation length scale of turbulence fluctuations. For example, there is a perturbation in the corona of 1-2\,Mm which causes an instantaneous position shift of radio sources with similar frequencies. If sources are separated by more than this length scale, the correlation disappears, as we show in Fig~\ref{fig:correlation}.

 

There has been much debate in recent literature on the length scales of the coronal turbulence that causes radio wave scattering. For example, \citet{1994Bastian} assume to this scale length to be the inner-scale of coronal turbulence, on the order of 1-10\,km in our case. However, \cite{2007Thejappa} theoretically determined this effective length scale to be $h = l^{1/3}_i l^{2/3}_o$ where $l_i$ and $l_o$ are the inner and outer scale respectively. They use empirical expressions for these scales, with $l_i = (r)^{\pm0.1}$ where $r$ is the heliocentric distance in unit of solar radii and $l_o = 8.82 \times 10^{-2} \bigg[\frac{R}{R_o}\bigg]^{0.82}$\,AU. Considering the heliocentric distance of 133.36\,MHz radio source to be 1.51 R\textsubscript{\(\odot\)} (according to 1 fold Saito density model), in our case these expressions lead to an effective scale length of $h\sim0.7$\,Mm, which is in the vicinity of our observationally determined scale length of 1-2\,Mm. This provides evidence that the random fine-scale motion of radio sources in our observed type II radio burst is due to density turbulence, and that the scale length of \cite{2007Thejappa} applies in our case.
Recent observations works \citep{2018_chen,Carley_2021} have found the density turbulence spectrum to be Kolmogorov like and \cite{Carley_2021} found the spatial scales of the turbulence spectrum to range from $0.2$\,Mm to $62$\,Mm placing our determined effective length scale in the inertial range of the turbulence.


\subsubsection{Excluding other systematic effects for the fine-scale motion} \label{ionosphere}
While analysing the fine scale motion of the type II sources we have to be sure if these source motions are due to a coronal phenomenon (turbulence in our hypothesis) and not due to ionospheric effects or imaging analysis artifacts. Below we discuss the reasons why we can exclude these effects with confidence.
\begin{enumerate}
        \item \textbf{Ionospheric effects:} 
        While traversing the Earth's ionosphere, low frequency radio sources can suffer apparent angular position offsets due to ionospheric propagation effects.
        It is conceivable that the observed correlated motion across nearby frequencies can arise due to such effects.
        However the timescale of the observed fluctuations, about 1\,s, is much faster than those associated with ionospheric variations (<50\,s)\citep{Jordan_2017}. 
        Also, the position offsets due to ionospheric refraction are inversely proportional to $f^{2}$ \citep[e.g.][]{Loi_2015}, where $f$ is the observing frequency.
        This implies that low frequency radio waves will be affected more than high frequencies and also provides a quantitative scaling which should hold for ionospheric propagation effects. 
        To check for this relationship, in Fig~\ref{fig:histogram} we show histogram plots of the instantaneous position shift magnitude ($\Delta s_{t+1} = s_{t+1}-s_{t}$) of sources for a single frequency from all six coarse channels. A Gaussian has been fitted to the histograms. The mean of the Gaussian model is shown by the black solid line and its $\sigma$ is also mentioned.
        The $1/f^2$ trend expected due to ionospheric propagation is not seen in either of these quantities, implying that they are not likely to arise due to ionospheric propagation effects.
    
     
         \item \textbf{Imaging artifacts:}
         During the process of imaging using self-calibration, it is possible for an arbitrary phase to be introduced which can then lead to position offsets in the image plane \citep{Pearson1984}. 
         For the analysis presented here, the self-calibration for each of the different frequencies was done only for a single time slice and the solutions obtained applied to the entire observing duration. In addition, self-calibration for each of the frequencies was done independently and the self-calibration based position shifts were corrected by moving the centroid of the self-calibrated image to that of the image obtained prior to self-calibration.
         More importantly, a single arbitrary shift in the image plane. arising due to self-calibration, can anyway not give rise to the time varying shifts seen in these data. 
         
         \item \textbf{An independent check:} 
         Out of an abundance of caution, we use the fortuitous presence of Virgo-A in the large MWA field-of-view to investigate the possible contamination from both these effects. 
         If the observed shifts correlated across nearby frequencies are a systematic effect arising out of either of the imaging process or ionospheric propagation, then a similar effect should also be seen for Virgo-A. 
         An independent Gaussian model was fit to Virgo-A for each time and frequency slice.
         The location of the peak of the Gaussian model was subjected to analysis identical to that done for the location of the type II source. 
         The position of Virgo-A was found to be very stable across time and frequency and no small scale motions correlated across neighbouring frequencies could be found.
         This provides an independent confirmation that the observed instantaneous position shifts in the type II emission are intrinsic and do not arise due to any ionospheric propagation effects or imaging artifacts.

\end{enumerate}
\section{Conclusions}
The first part of this work focused on the understanding of how shocks can cause split bands in the harmonic band of a particular type II event using imaging spectroscopy of MWA and EUV observations of AIA/SDO. Imaging the radio sources from these sub-bands revealed that the sources are separated in space by up to $\sim$372\,Mm in the plane-of-sky. The radio emissions are likely generated in different coronal loops ahead of the shock front, generating plasma emission at different frequencies during the burst.
This provides rare imaging evidence against the popular interpretation that the band-splitting is caused by emission upstream and downstream of the shock front and calls into question the widespread use of this interpretation to calculate important properties such as density jump across the shock, Mach number, and magnetic field. Our study demonstrates the importance of imaging for verifying which of these competing theories are behind the band-splitting in type IIs.

In the second part of the work we have used imaging-spectroscopy to study the effect of small scale perturbations to fine scale motion of type II sources. At low radio frequencies, these perturbations are believed to arise due to coronal density inhomogeneities which the radiation encounters while propagating out through the corona. We observed that the fine-scale motion of the sources exhibited a systematic correlation across nearby frequencies. To investigate this we developed a new technique of cross correlation of instantaneous change in direction and magnitude of the radio sources of nearby frequency channels (dot-product correlation). We interpret such correlations as arising due to small scale density perturbations from coronal turbulence. These perturbations lead to similar displacement of sources at nearby frequencies, spaced up to 0.5\,MHz apart and corresponds to a spatial separation of $\sim1- 2$\,Mm at coronal heights of interest. If this is interpreted as the effective length scale of turbulence, our observational work agrees with the theoretical findings of \citep{2007Thejappa}.

Hence, our works provide a unique way to directly characterize the turbulence in different regions of shocks and the corona. One can also potentially obtain the intensity of these fine structures for a large range of frequencies and get the power density spectra, giving insight into the spectral characteristics of coronal turbulence \citep{2018_chen,Carley_2021}. Such studies can provide the range of length scales of the coronal density perturbations and provide an opportunity to check if the determined length scale lies in the inertial range. This will be an interesting avenue to explore in future studies.

\begin{acknowledgements}
 
This scientific work makes use of the Murchison Radio-astronomy Observatory (MRO), operated by the Commonwealth Scientific and Industrial Research Organisation (CSIRO).
We acknowledge the Wajarri Yamatji people as the traditional owners of the Observatory site. Support for the operation of the MWA is provided by the Australian Government's National Collaborative Research Infrastructure Strategy (NCRIS), under a contract to Curtin University administered by Astronomy Australia Limited. We acknowledge the Pawsey Supercomputing Centre, which is supported by the Western Australian and Australian Governments. 
SB is supported by the Hamilton PhD Scholarship at DIAS. EPC is supported by the Schr\"odinger Fellowship at DIAS.
DO acknowledges support of the Department of Atomic Energy, Government of India, under the project no. 12-R\&D-TFR-5.02-0700. 
We would like to thank Dr. Peijin Zhang for his valuable discussions.
This research has also made use of NASA's Astrophysics Data System (ADS). 
This research has made use of the NASA/IPAC Extragalactic Database (NED), which is operated by the Jet Propulsion Laboratory, California Institute of Technology, under contract with the National Aeronautics and Space Administration. The authors would like to acknowledge AIA, GOES and Learmonth team for open access to their data.We thank the referee for the helpful comments to improve the paper.

\end{acknowledgements}

%
%
\bibliography{aanda.bib}{}
\bibliographystyle{aa}

\end{document}